\documentclass{ws-mpla}
\newcommand{\no}{\nonumber}

\newcommand{\eea}{\end{eqnarray}}
\newcommand{\bea}{\begin{eqnarray}}

\begin{document}

\markboth{Authors' Names}
{Instructions for Typing Manuscripts (Paper's Title)}

%
\catchline{}{}{}{}{}
%

\title{UNITARITY OF NONCOMMUTATIVE FIELD THEORIES\\
FROM STRING THEORY\footnote{Based on the talk given at the conference: ``Spacetime and Fundamental Interactions: Quantum Aspects. A conference to honour A. P. Balachandran's 65th birthday'', Vietri, 26-31 May 2003}}

\author{\footnotesize ALESSANDRO TORRIELLI}

\address{Dipartimento di Fisica ``Galileo Galilei'', Universit\`a di Padova\\
INFN Sezione di Padova\\
via F. Marzolo 8, 35131 Padova, Italy\\
torriell@pd.infn.it}

\maketitle

\begin{abstract}
We improve the study of the lack of perturbative unitarity of noncommutative space-time quantum field theories derived from open string theory in electric backgrounds, enforcing the universality of the mechanism by which a tachyonic branch cut appears when the Seiberg-Witten limit freezes the string in an unstable vacuum. The main example is realized in the context of the on-shell four-tachyon amplitude of the bosonic string, and the dependence of the phenomenon on the brane-worldvolume dimension is analysed. We discuss the possibility of a proof in superstring theory, and finally mention the NCOS limit in this framework.

\end{abstract}

\ccode{PACS 11.25.Db, 11.10.Kk, 11.55.Bq. \ \ \ \ \ \ \ \ \ \ \ No DFPD 03/TH 27}

\section{Introduction}	

Since the seminal work of Seiberg and Witten\cite{sw} it was realized that open strings in the presence of an antisymmetric constant background are effectively described at low energy by certain noncommutative field theories, identified by a precise set of Feynman rules\cite{f}. These theories are similar to the ones appearing in previous works\cite{dfr}, where the necessity of a space-time uncertainty principle when quantum mechanics couples with general relativity was realized, and subsequently a consistent quantum field theory in such a noncommutative setup was tentatively constructed. However, there are many important differences among these two formulations which have some drastic effects, in particular for what concerns the problem of unitarity: the former were proven to lose perturbative unitarity when time is involved in the noncommutativity\cite{gm,bgnv}, while the latter are made free from this problem\cite{bdfp,ls2}. We will not discuss here the problems related to the latter, and to the extension of the quantum field theory axiomatic framework to the noncommutative case (see also \cite{av}). We will instead concentrate our attention on the former. 

The derivation by Seiberg and Witten was originally made for magnetic backgrounds only, corresponding to space-space noncommutativity (for the light-like case see \cite{agm}). For electric backgrounds it is well known that problems are present when the electric field approaches a critical value $E_{cr}$, beyond which the string develops a classical instability\cite{b,bp,sst1}: tachyonic masses appear in the spectrum both for neutral (which is the case we will consider here) and charged open strings, connected with the vanishing of the effective tension and an uncontrolled growth of the oscillation amplitude of the modes in the direction parallel to the field. For purely charged strings this phenomenon (which has no analog in particle field theory) coexists with the quantum instability due to pair production which is the analog of the Schwinger phenomenon in particle electrodynamics\footnote{As a note, we recall that for neutral strings ($+q$-charge on one end, $-q$ on the other), one has $E_{cr}=1/(2\pi \alpha' |q|)$. For charges $q_1\neq q_2$ on the two boundaries, one finds that the pair production rate diverges at the same critical value of the classical instability $E_{cr} = 1/(2\pi \alpha' |max q_i |)$}. When one tries to perform the Seiberg-Witten zero-slope limit to noncommutative field theories one reaches a point in which the ratio between the electric field and its critical value becomes greater than one, and the string enters the region of instability. We will show that precisely at this point a flip-mechanism produces the appearance of the tachyonic branch cut due to the closed string sector in the non-planar diagrams, which is responsible for the lack of unitarity of the limit amplitude\footnote{Space-time noncommutative field theories present besides a series of exotic behaviours\cite{sst2,br}}. We relate the breaking of perturbative unitarity of these space-time noncommutative theories with the fact that they coincide with those obtained by freezing string theory in an unstable vacuum. This vacuum is likely to decay in the full string theory to a suitable configuration of branes. 

Results will be presented as follows: in section 2 we will describe the generic bosonic string amplitude for tachyons on external legs, and then specify it for the two-point case, for which we show the appearance of the flip mechanism\cite{mio}. This will be a very primitive setup, which in its elementary form already reveals the phenomenon. We will then perform some improvements through a more complicated example: in section 3 we will analyse the four-tachyon amplitude, in order to overcome the problem of the off-shell continuation. We will then stress the universality of the phenomenon with respect to the dimension of the brane-worldvolume\footnote{An analysis for vector amplitudes is contained in \cite{roberto}}. In section 4 we will mention the NCOS limit. Finally we will conclude and comment on future improvements, especially the perspective to reproduce this mechanism starting from a fully consistent superstring theory.

\section{General Setup and Two-point Tachyon Amplitude}

We take as action for a bosonic open string attached to a D-brane lying in the first $p+1$ dimensions with an antisymmetric constant background on its worldvolume the following
\begin{eqnarray}
\label{1}
S={{1}\over{4\pi {\alpha}'}}\int_{C_2}d^2 z \, (g_{ij} {\partial}_a X^i {\partial}^a X^j - 2 i \pi {\alpha}' B_{ij} {\epsilon}^{ab}{\partial}_a X^i {\partial}_b X^j).     
\end{eqnarray}
At one loop the string world-sheet is the cylinder $C_2 = \{0\leq \Re w\leq 1, w \equiv  w + 2 i \tau$\}; $\tau$ is the modulus. We are taking the indices $i$ and $j$ to live on the brane. 

The one loop propagator we adopt in this situation\footnote{The reader is referred to the literature\cite{bcr,gkmrs,kl,lm,crs}}, with the new boundary conditions imposed by the $B$-term, can be found in \cite{ad}. If one sets $w=x+iy$, the relevant propagator on the boundary of the cilinder ($x=0,1$) can be written as
\begin{eqnarray}
\label{2}
G(y,y')={{1}\over{2}}{\alpha}' g^{-1} \log q - 2 {\alpha}' G^{-1} \log \Big[{{q^{{1}\over {4}}}\over {D(\tau)}} \, {\vartheta_4}({{|y-y'|}\over {2 \tau}}, {{i}\over {\tau}})\Big],\, \, \, x\neq x', 
\end{eqnarray}
\begin{eqnarray}
\label{3}
G(y,y')={{\pm i \theta }\over{2}} {\epsilon}_{\perp} (y-y') - 2{\alpha}' G^{-1} \log \Big[{{1}\over {D(\tau)}} {\vartheta_1}({{|y-y'|}\over {2 \tau}}, {{i}\over {\tau}})\Big],\, \, \, x=x',
\end{eqnarray}  
where $q=e^{-{{\pi}\over{\tau}}}$, $\pm$ correspond to $x=1$ and $x=0$ respectively, and ${\epsilon}_{\perp} (y) = sign(y) - {{y}\over{\tau}}$.
The open string parameters are:
\begin{eqnarray}
\label{4}
G=(g-2\pi {\alpha}' B)g^{-1} (g+2\pi {\alpha}' B)
\end{eqnarray}
which is the open string metric, and
\begin{eqnarray}
\label{5}
\theta=-{(2\pi{\alpha}' )}^2 {(g+2\pi {\alpha}' B)}^{-1}B {(g-2\pi {\alpha}' B)}^{-1}
\end{eqnarray}
which is the noncommutativity parameter. ${\vartheta_{1,4}}(\nu, \tau )$ are Jacobi theta functions, while $D(\tau)={\tau}^{-1} {[\eta({i\over \tau})]}^3$, where $\eta$ is the Dedekind eta function\cite{p}.
With this propagator and the suitable modular measure, the amplitude for the insertion of $N$ tachyonic vertex operators at $x=1$ and $M-N$ at $x=0$ turns out to be\cite{ad}: 
\begin{eqnarray}
\label{6}
A_{N.M}&=&{\cal{N}}{(2\pi)}^{d} {({\alpha}')}^{\Delta} {G_s}^M \int_0^{\infty} {{d\tau}\over{\tau}} {\tau}^{-{{d}\over{2}}} {[\eta(i\tau)]}^{2-d} q^{{{1}\over{2}}{\alpha}' Kg^{-1} K}  \nonumber \\
&&\times \Big(\prod_{a=1}^M \int_0^{y_{a-1}} dy_a \Big) \prod_{i=1}^N \prod_{j=N+1}^M {\Bigg[ q^{{1}\over {4}} \, {\vartheta_4}({{|y_i - y_j|}\over {2 \tau}}, {{i}\over {\tau}}) / D(\tau)\Bigg]}^{2 {\alpha}' k_i G^{-1} k_j } \no \\
&&\times \prod_{i<j=1}^N e^{- {{1}\over{2}}i {\epsilon}_{\perp} (y_i - y_j)k_i \theta k_j} {\Bigg[ {\vartheta_1}({{|y_i -y_j|}\over {2 \tau}}, {{i}\over {\tau}}) / D(\tau)\ \Bigg]}^{2 {\alpha}' k_i G^{-1} k_j }\nonumber \\
&&\times \prod_{i<j=N+1}^M e^{{{1}\over{2}}i {\epsilon}_{\perp} (y_i - y_j)k_i \theta k_j} {\Bigg[ {\vartheta_1}({{|y_i -y_j|}\over {2 \tau}}, {{i}\over {\tau}}) / D(\tau)\ \Bigg]}^{2 {\alpha}' k_i G^{-1} k_j }.
\end{eqnarray}
Here ${\cal{N}}$ is the normalization constant, $d=p+1$, $\Delta = M {{d-2}\over{4}} - {{d}\over {2}}$, $G_s$ is the open string coupling constant, $K=\sum_{i=1}^N k_i$ is the sum of all momenta associated with the vertex operators inserted on the $x=1$ boundary, and $y_0 = 2\tau$. We will consider here always a fixed cyclic order of the vertex insertions. Whenever $N\neq0$ and $M>N$, this amplitude corresponds to non-planar graphs, which in the field theory limit will become the non-planar contribution to the noncommutative amplitude. We have also omitted a global delta function due to momentum conservation, and the traces of the Chan-Paton matrices.

By rescaling $t=2\pi {\alpha}' \tau$ and $\nu_{1,2} ={{y_{1,2}}/ {2 \tau }}$, and setting $\nu_{2} =0$ to fix the residual invariance, the non-planar two-point function can be written as follows: 
\begin{eqnarray}
\label{7}
A_{1.2}&=&{\cal{N}} {G_s}^2 \, 2^{{3 d} \over 2}{\pi}^{{{3 d}\over{2}}-2} {\alpha '}^{{{d}\over{2}}-3} \int_0^{\infty} dt \, t^{1-{{d}\over{2}}} \, {\Big[ \eta({{i t}\over{2\pi {\alpha '}}})\Big] }^{2-d}\times \nonumber \\
&&\, e^{- {{{\pi}^2}{\alpha '}^2 \over {t}} kg^{-1} k} 
\int_0^1 d\nu {\Bigg[{{e^{-{{{\pi}^2}{\alpha '}\over{2t}}} {\vartheta_4}(\nu, {{2 \pi i {\alpha '}}\over {t}})} \over { {{2 \pi{\alpha '}}\over{t}} {[\eta({2\pi i {\alpha '}\over {t}})]}^3 }}\Bigg]}^{- 2 {\alpha '} k G^{-1} k}.
\end{eqnarray}
This is a form suitable for the field theory limit. We perform the zero-slope limit of Seiberg and Witten, which consists in sending ${\alpha}' \to 0$ keeping $\theta$ and $G$ fixed (in this case we will keep $t$ and $\nu$ fixed as well). This can be done setting ${\alpha}' \sim {\epsilon}^{1\over 2}$ and the closed string metric $g\sim \epsilon$, and then sending $\epsilon \to 0$ \cite{sw}. One obtains \cite{ad}
\begin{eqnarray}
\label{8}
A_{1.2}^{lim} ={\cal{N}} 2^{{3 d} \over 2}{\pi}^{{{3 d}\over{2}}-2} {g_f}^2 \int_0^{\infty} dt t^{1-{{d}\over{2}}} e^{- m^2 t \, + \, k\theta G\theta k/4 t}\int_0^1 d\nu e^{- t\, \nu (1 - \nu )\, k G^{-1} k}.
\end{eqnarray}
This reproduces the expression for the two-point function in the noncommutative $\phi^3$ theory\cite{dn,s,tesi} with coupling constant $g_f = G_s \, {{\alpha}'}^{{{d-6}\over {4}}}$. We choose now $d=2$. This case is peculiar, since the tachyon mass ${m^2} = {{2-d}\over {24 {\alpha}'}}$ goes to zero. In two dimensions the background is an electric one and noncommutativity is necessarily space-time. 

In order to interpret the field theory analysis, one continues the expression (\ref{7}) in the complex variable $k^2=k G^{-1} k$. It has a branch cut driven by the small-$t$ (closed-channel) behavior 
\begin{eqnarray}
\label{bigt}
\exp \Big( - k^2 \, {{{\pi}^2 {{\alpha '}^2 \, [1 - {(E / E_{cr})}^2 ]}}\over {t}}\Big).
\end{eqnarray}
The field-theory limit retains in turn another branch cut $Re [S] = Re [- k^2] > 0$ from the large-$t$ behaviour
\begin{eqnarray}
\label{smallt}
\exp \Big(- k^2 \, t \, \nu (1 - \nu)\Big).
\end{eqnarray}
$E_{cr} ={{g}\over{2 \pi {\alpha '}}}$ is the critical value of the electric field\footnote{We have taken unitary charges (see (\ref{1})) and a closed string metric $g_{\mu \nu } = g \, {\eta }_{\mu \nu }$}. The small-$t$ cut is on the physical side $Re [S] = Re [- k^2] > 0$. But we see that it is driven by a quantity which changes sign if the electric field overcomes its critical value, that is when the string enters the classical instability region. This is exactly what the Seiberg-Witten limit produces, since it scales ${(E / E_{cr})}^2 \sim {(\alpha'/g)}^2 \sim 1 / \epsilon$. In the field theory limit the branch cut flips to the unphysical side\cite{mio}, and in fact the amplitude (\ref{8}) for $m^2 = 0$ has two branch cuts, one of which is physical, the other one tachyonic, coalescing at the origin. By treating $m^2$ as a free parameter independent of $d$ and by taking it positive\footnote{This is the point of view of many works by Di Vecchia et al.\cite{div1,div2} in deriving field theory amplitudes from the bosonic string, and it is also the approach usually adopted in the noncommutative case}, one can see the shift of (\ref{bigt}) to a cut starting from $S > 4 m^2$ which gives the physical pair-production cut, and a closed branch cut  starting at $S > 4 m^2 / [1 - {({{E}/{E_{cr}}})}^2]$, which flips in the limit.   

\section{The four-point Tachyon Amplitude}

In this section we analyse the non-planar amplitude (\ref{6}) with four external tachyons $A_{2.4}$. As anticipated, this allows us to set ${k_n}^2  = - m^2$ on mass-shell. We define the Mandelstam variables\footnote{Indices are raised and lowered with the open string metric (also denoted by a dot in the following) where not differently specified} as $S=-(k_1 + k_2)^2$, $T=-(k_1 + k_4)^2$, $U=-(k_1 + k_3)^2$, and study the behaviour of the amplitude with respect to the variable $S$, setting the other ones at suitable physical values\cite{roberto}. We choose the gauge $\nu_4 = 0$. 
\begin{eqnarray} \label{ampl24tach2} 
  A_{2.4} &=& \mathcal{N} {(2 \pi)}^d g_f^4
              \int_0^\infty dt
	      t^{3-d/2} {{[\eta(\frac{it}{2\pi\alpha'})]}^{2-d}} 
	      {e^{-{\frac{\pi^2{\alpha'}^2}{t}} K g^{-1} K}}
	      \left[ \prod_{n=1}^3 \int_0^{\nu_{n-1}}d\nu_n \right]  
 	      \nonumber \\
	  && \times   
              e^{-\frac{i}{2}k_1\theta k_2 (1 - 2\nu_{12}) 
	      +\frac{i}{2}k_3\theta k_4 (1 - 2 \nu_3)}
	      \prod_{n=1,2} \prod_{m=3,4}
	      \left[\frac{e^{-\frac{\pi^2\alpha'}{2t}} 
	      \vartheta_4(\nu_{nm},\frac{2\pi i\alpha'}{t})}
              {(\frac{2\pi\alpha'}{t})
	      [\eta(\frac{2\pi i\alpha'}{t})]^3} \right]^{2\alpha'
		k_n \cdot k_m}  \nonumber \\
          && \times  
	      \left[\frac{\vartheta_1
	      (\nu_{12},\frac{2\pi i\alpha'}{t})}
	      {(\frac{2\pi\alpha'}{t})
	      [\eta(\frac{2\pi i\alpha'}{t})]^3} \right]^{2\alpha'
	      k_1 \cdot k_2}  \left[\frac{\vartheta_1
	      (\nu_{3},\frac{2\pi i\alpha'}{t})}
	      {(\frac{2\pi\alpha'}{t})
	      [\eta(\frac{2\pi i\alpha'}{t})]^3} \right]^{2\alpha'
	      k_3 \cdot k_4} 
\end{eqnarray}
where $\nu_{mn} = \nu_m - \nu_n$.  With such a scattering, we should find another trick in order to avoid the tachyonic mass rather than going to brane-worldvolume dimension two. We will again treat the mass parameter as an independent positive quantity, which will reproduce the field theory result with a positive mass. By using momentum conservation, the on-shell condition and the asymptotic form of the special functions contained in the integrand, we derive the branch cut retained by the field-theory in the large-$t$ behaviour
\begin{eqnarray}
\label{large4}
t^{3-d/2} \, e^{-m^2t}e^{t\sum_{n<m=1}^{4}\nu_{nm}(1-\nu_{nm}) k_n \cdot k_m},
\end{eqnarray}
which is\cite{gm,roberto} $S>4m^2$. 

The small-$t$ behaviour reads
\begin{eqnarray}
\label{small4}
t^{7/3 - d/6}
\, \, e^{-\frac{\pi^2{\alpha'}^2}{t}[\frac{2-d}{6\alpha'}+K g^{-1}K]},
\end{eqnarray}
where we recall $K=k_1 + k_2$. First of all, we notice that in $d=26$ this produces the usual closed-string tachyonic pole. In different dimensions it produces instead a branch cut\cite{gsw}, driven again by the ratio of the electric field over its critical value\footnote{We still choose the field of electric type, turned on in the $0-1$ directions, and for simplicity disregard the contribution of $k_{\perp}$}: for a stable string the cut is for    
\begin{eqnarray}
\label{massachiusa}
S \, > \, 4 m^2 / [1 - {(E/E_{cr})}^2 ] \, = M^2,
\end{eqnarray}
where $M^2$ is the mass of the closed tachyon. This branch cut flips in the limit, when the electric field overcomes its critical value\cite{roberto}.

One may ask what happens in the critical dimension, and if the field theory will present a pole. This is not true, in $d=26$ the field theory limit will have a tachyonic branch cut. One can realize it by looking at the field theory amplitude 
\begin{eqnarray}
A_{2.4}^{lim} &=& g_f^4 \, \mathcal{N} {(2 \pi )}^d 
              e^{-\frac{i}{2}k_1\theta k_2}
	      e^{\frac{i}{2}k_3\theta k_4}\int_0^\infty dt
	      \, t^{3-d/2} e^{-m^2 t} 
	      e^{-\frac{1}{4t} 
	      K \theta^2 K}  \nonumber \\
	  &\times  & \left( \prod_{n=1}^3 \int_0^{\nu_{n-1}}d\nu_n \right)
	      e^{i(\nu_{12})k_1\theta k_2}
	      e^{-i\nu_3k_3\theta k_4}\prod_{n<m=1}^4
	      e^{t \nu_{nm}(1-\nu_{nm})k_n \cdot k_m} \nonumber 
\end{eqnarray}
whose integral over $t$ gives a Bessel function. There can be an intuitive explanation of this fact\cite{abz}: the tower of closed poles present in $d=26$, among which the above tachyon pole is the first one, is easily seen to be separated by a spacing ${[\alpha' |1 - {(E / E_{cr} )}^2 |]}^{- 1} \sim \sqrt{\epsilon}$. This spacing goes to zero and the poles cumulate in the limit, producing a continuum cut.

\section{NCOS}

Consider the amplitude (\ref{ampl24tach2}) of the previous section: if one performs a different limit on the string parameters than the Seiberg-Witten one, namely
\begin{eqnarray}
\label{ncoslim}
\alpha' \sim const. \qquad , \qquad E \sim 1 / \epsilon \qquad , \qquad g \sim 1 / \epsilon 
\end{eqnarray}
in such a way that $G,\theta \sim const.$ and $E/E_{cr} \sim 1 - \epsilon$, then the amplitude becomes
\begin{eqnarray} \label{ampl24tach2ncos} 
  A_{2.4} &=& \mathcal{N} {(2 \pi)}^d g_f^4
              \int_0^\infty dt
	      t^{3-d/2} {{[\eta(\frac{it}{2\pi\alpha'})]}^{2-d}} 
	      {e^{-{\frac{\pi^2{\alpha'}^2}{t}} K g_{(lim)}^{-1} K}}
	      \left[ \prod_{n=1}^3 \int_0^{\nu_{n-1}}d\nu_n \right]                          \nonumber \\
	  && \times   
              e^{-\frac{i}{2}k_1\theta k_2
	      +\frac{i}{2}k_3\theta k_4
	      +i\nu_{12}k_1\theta k_2
	      -i\nu_3k_3\theta k_4}
	      \prod_{{n=1,2} \over {m=3,4}}
	      \left[\frac{e^{-\frac{\pi^2\alpha'}{2t}} 
	      \vartheta_4(\nu_{nm},\frac{2\pi i\alpha'}{t})}
              {(\frac{2\pi\alpha'}{t})
	      [\eta(\frac{2\pi i\alpha'}{t})]^3} \right]^{2\alpha'
		k_n \cdot k_m}  \nonumber \\
          && \times  
	      \left[\frac{\vartheta_1
	      (\nu_{12},\frac{2\pi i\alpha'}{t})}
	      {(\frac{2\pi\alpha'}{t})
	      [\eta(\frac{2\pi i\alpha'}{t})]^3} \right]^{2\alpha'
	      k_1 \cdot k_2}  \left[\frac{\vartheta_1
	      (\nu_{3},\frac{2\pi i\alpha'}{t})}
	      {(\frac{2\pi\alpha'}{t})
	      [\eta(\frac{2\pi i\alpha'}{t})]^3} \right]^{2\alpha'
	      k_3 \cdot k_4}.  
\end{eqnarray}
We choose $d=26$ in what follows. Since $\alpha'$ remains constant, this is a string amplitude which contains an infinite tower of contributions from the string spectrum, and is not a field theory amplitude. The electric field approaches indefinitely its critical value from below, never passing the stability threshold. The only difference with respect to the case before the limit is in the exponential which contains the closed string metric $g$, whose entries in the $0-1$ block tend to zero. The large-$t$ behaviour is the same as in the previous section; the small-$t$ reduces again to (\ref{small4}), but the result of the integration now gives $1/{[\frac{2-d}{6\alpha'}+K_{\perp}^2]}$, where $K_{\perp}^2$ is positive definite. We see therefore that the amplitude is protected against the flip-mechanism. The closed string pole disappears, which points towards the decoupling of the closed string sector. We roughly reproduce at this perturbative order in the bosonic string theory the Noncommutative Open String (NCOS) limit\cite{sst1,gmms}: this is an open string theory limit decoupled from closed string.\\ \section{Conclusions}
We have shown that it is generally possible to connect the lack of unitarity of string-derived noncommutative field theories with the appearance of a tachyonic branch cut due to the closed sector in non-planar diagrams, which appears when one enters the region of instability of the neutral open string in the background of an electric field, when it overcomes a critical value basically equal to the string tension over the endpoint charge. This phenomenon was seen in the calculation of the bosonic two-point function, where an off-shell continuation is compulsory in order to compare with the field-theoretical analysis. We removed this arbitrariness by considering an on-shell four-tachyon amplitude and we realized the universality of this process. We discussed the dependence on the worldvolume of the brane and what happens for a brane which fills the space.\\ 
The next step should be to single out this phenomenon in a fully consistent superstring theory, where the presence of the tachyon contribution is ruled out from the beginning. A typical setup could be for example type IIB superstring with a stack of coincident $D3$-branes, with an NS-NS two-form background turned on on their worldvolume, and analyzing incoming massless states from the spectrum of the open string living on them. The effective field theory action will be in this case a space-time noncommutative supersymmetric theory. From this kind of analysis one could also try to understand some issues regarding UV/IR singularities which have obscure meaning on the field theory side even in the magnetic case\cite{mvs,bgnv,rr,clz}. A more ambitious project regards the formulation of this analysis in the framework of the open string field theory. The aim should be to reproduce not only the complete analysis of singularities for off-shell amplitudes, but to clarify the fate of the unstable vacuum and indicate a path of brane-decay for this situation.\\       
Finally, we mentioned a bosonic realization of the NCOS regime, where open strings decouple from closed ones and from gravity. This theory is quite particular, since it contains open strings only, and it may help in singling out some features of pure open strings and gauge theories when slowly turning off the gravity sector\cite{amss,kp}. \\ \section*{Acknowledgments}
This work has been done in collaboration with Antonio Bassetto and Roberto Valandro of the University of Padua. I wish to thank also L. Bonora for discussions and for drawing our attention to the NCOS limit, and R. Russo for useful discussions, and C.-S. Chu, J. Gomis, S. Kar and M. Tonin for suggestions.

\end{document}